# Sociofísica en el caso mexicano

Ignacio Ibarra López[1], Hugo Hernández-Saldaña[2]
[1]Instituto de Ciencias de Gobierno y Desarrollo Estratégico, Benemérita Universidad Autónoma de Puebla, Puebla Pue., México
[2] Departamento de Ciencias Básicas, Universidad Autónoma Metropolitana en Azcapotzalco, México D. F., México
Teléfono (55) 5318-9575    E-mail: hhs@correo.azc.uam.mx

*Resumen* — **Con la disponibilidad de datos electorales en formato electrónico los físicos y matemáticos iniciaron sus propios análisis de los datos. Más allá de los modelos teóricos desarrollados para los procesos electorales, nuevas regularidades empezaron a surgir de los datos en muchas elecciones y en muchos países. Aquí discutimos algunas de las regularidades estadísticas encontradas para México. En particular, la aparición de leyes de potencia y modelos Margarita (*Daisy models*) que se pueden asociar al voto corporativo. Los métodos utilizados son de estadística y física estadística estándar. Como una aplicación particular, discutimos una aproximación entrópica a la distribución de votos del Partido Revolucionario Institucional y demostramos que éste método NO permite diferenciar una compra masiva de votos pues su comportamiento ha sido sorprendentemente consistente. De igual forma otros métodos han resultado infructuosos en determinar signos de fraude.**

*Palabras Clave* – **sociofísica, análisis forense de elecciones, estadística**

*Abstract* — **The availability of electoral data in electronic format allowed to physicist and mathematicians their own analysis. Beyond theoretical models for electoral processes, new regularities had been found in the data analysis of many countries and many elections. Here, we discuss some of the statistical regularities found for México, from power laws Daisy models linked in corporate vote. The methods used are those from standard statistics and statistical mechanics. As a particular application we discuss an entropic approach to the distribution of votes for the Partido Revolucionario Institucional and we demonstrate that this kind of method DOES NOT allow to discover a massive buy of vote. The argument is that such a distribution has been amazingly consistent in the last two decades elections. Other methods have been unsuccessful to discover signal of fraud.**

I. INTRODUCCIÓN

El origen de la curiosidad de parte de los físicos en este tipo de problemas radica en la evidencia de que el comportamiento colectivo de muchos individuos puede ser modelado o entendido como si se trataran de partículas interactuantes. Esto es, partículas con alguna dinámica (posiblemente desconocida) cuyas propiedades pueden ser medidas y modeladas a través de la "mecánica estadística" [1-5]. Por supuesto que la dinámica puede ser bastante complicada tomando en consideración la existencia de interacciones "psicológicas" o "culturales" entre los individuos de las cuales apenas tenemos vistazos cuantitativos. Sin embargo, el acceso a datos en formato electrónico así como los estudios y mediciones en tiempo real (e.g. encuestas de salida y conteos rápidos) ha dado un impulso a su estudio. El estudio, por parte de físicos, de datos *empíricos* en elecciones es un fenómeno reciente, lo cual explica por qué aunque las referencias van en aumento, aun son escasas comparadas con otras áreas [6-22]. El caso mexicano no es la excepción [23-28].

A través de los datos proporcionados por las autoridades electorales mexicanas se ha podido identificar algunas de las propiedades estadísticas de una votación desde un punto de vista diferente al de los científicos sociales.

Por ejemplo, se han encontrado algunas propiedades de las tasas de participación de votantes (*voter turnou*t)[21] y del tipo de distribución del voto a favor de algunos de los partidos participantes[25]. Dentro de los esfuerzos académicos destaca la iniciativa de probar estadísticamente la existencia de fraude en elecciones presidenciales [29,30] o tratar de entender dichos procesos[31-34].

En este trabajo presentamos un resumen de los hallazgos hechos en este campo en relación al caso mexicano, en lo que se ha dado a conocer como sociofísica. Es decir, no sólo el estudio forense de las elecciones (*election forensics*), sino la búsqueda de modelos que expliquen el comportamiento de los electores. Un punto importante en la perspectiva considerada en este trabajo es su utilización en la detección de fraudes, que como veremos, ha resultado infructuoso [26,28,35].

II. METODOLOGÍA

*A. Construcción de los Datos*

Los datos electorales son de libre acceso y esto está garantizado por las leyes de transparencia que rigen actualmente en México. Estos datos pueden obtenerse en la sección de estadísticas de la página http://www.ife.org.mx.[36] Ahí se pueden descargar los datos de las elecciones históricas federales desde 1994 hasta la actual a nivel casilla para los resultados del Conteo Distrital, i.e., los resultados oficiales y con implicaciones jurídicas. Esta base de datos permite el acceso de resultados a nivel de casillas. Además, existe acceso a los datos del PREP (Programa de Resultados Electorales Previos) de las elecciones presidenciales [37]. Los casos de las elecciones federales intermedias y locales son más difíciles de localizar, pero también son de libre acceso y se pueden solicitar de manera expresa en la página de transparencia del mismo Instituto [38].

Los formatos en los que vienen capturados los datos varían de elección en elección y es importante tener cuidado al realizar la limpieza de los mismos. Uno de los problemas para los archivos más antiguos es que los campos que están en blanco en las actas aparecen vacíos en el reporte y no son diferenciados de otros campos vacíos del sistema. Por ejemplo los campos para comentarios.

En un afán de hacer la estadística y el tratamiento de los datos de la manera más transparente posible se está explorando la posibilidad de usar el programa computacional PANDAS [39]. Lo anterior con la finalidad de asegurar la reproducibilidad de los datos y de los análisis. El uso de R también se ha explorado [40], aún cuando los análisis que se han realizado hasta el momento han sido hechos usando programas computacionales en Fortran y graficadores de software libre(gnuplot y xmgrace).

B. Temas no tratados

Desde hace algunas elecciones las actas de cada casilla ha sido capturada y permite un análisis muy pormenorizado, como el realizado por A. Crespo y sus colaboradores [33] para las elecciones presidenciales del año 2006. Todos esos registros, es decir la copia electrónica de las actas de casilla, son de acceso público y pueden solicitarse al IFE en su totalidad. En el presente trabajo no revisamos dicha documentación pues estamos interesados por el momento, en propiedades generales y de carácter estadístico. Como puede verse y explicaremos más adelante, la búsqueda de propiedades estadísticas universales o de cierto grado de independencia de la sociedad y del tiempo de la elección está en la mira de todos los que realizamos este tipo de estudios.

Otro tema que hemos dejado fuera de la discusión es el caso de la existencia de un algoritmo durante la elección presidencial de 2006 y otras discusiones sobre dicha elección, un resumen sobre esos puntos de vista se encuentra en la referencia [30].

C. Hipótesis

1. Hipótesis nula

Una cuestión fundamental en todo estudio de carácter estadístico y, en particular, uno que puede ser tan elusivo como el de los procesos electorales es el evitar la formulación y la prueba de hipótesis a partir de un solo conjunto de datos. Siguiendo a Good página 3[41], se plantea que:

"Las fuentes de error al aplicar procedimientos estadísticos son legión e incluyen todo lo siguiente:

* Usar el mismo conjunto de datos para formular hipótesis y entonces probar estas hipótesis.
..."

De tal manera que todas las propiedades estadísticas que se han analizado por nosotros o por otros han de verificarse en varias elecciones. Ello incluye todas las elecciones o bases de datos más o menos equivalentes y que sean accesibles.

Otros errores comunes en la aplicación de análisis estadísticos que se mencionan son:

"Tomar muestras de la población incorrecta o fallar en la especificación previa de la población o poblaciones sobre las cuales las inferencias son hechas."
"Fallar en describir muestras que sean al azar y representativas".

Los casos anteriores son relevantes pues: "De hecho, cualquier grupo o agregación (*cluster*) de individuos que viven, trabajan, estudian u oran juntos pueden fallar en ser representativos por cualquiera o por todas de las siguientes razones[42]:
1.- Compartieron la exposición al mismo ambiente físico o social.
2.- Auto-selección en pertenecer al grupo.
3.- Compartir comportamientos, ideas o enfermedades entre los miembros del grupo."

De tal manera que cualquier estudio debe evitar estos errores y los subsiguientes enumerados en la referencia [41].

III. RESULTADOS

A. El ordenamiento temporal del PREP-presidencial 2000-2012 y la dinámica en la generación de la información.

Uno de los procesos que ha generado un gran debate es la fiabilidad de los órganos electorales mexicanos. En particular, para la elección presidencial de 2006 [43]. Uno de los principales cuestionamientos, fue la existencia de un fraude en contra del candidato de la coalición de partidos de izquierda. En esta elección se presentaron tres candidatos: Felipe de Jesús Calderón Hinojosa (FJCH) del Partido Acción Nacional (PAN), Roberto Madrazo Pintado (RMP representante del Partido Revolucionario Institucional (PRI) y del Verde Ecologista de México (PVEM) y Andrés Manuel López Obrador (AMLO), quien encabezaba una coalición de partidos de izquierda. Los resultados que aparecieron en la página del Programa de Resultados Electorales Preliminares (PREP) durante esa jornada fueron controversiales debido a que presentaron un cruce entre FJCH y AMLO lo cual sugería la posible manipulación de los resultados. A continuación demostraremos que con un análisis a profundidad de la elección este argumento no se sostiene. En la Fig. 1 mostramos los datos capturados durante esa jornada por Baqueiro[44] tomando el porcentaje de votos obtenidos por los tres candidatos principales, como función del porcentaje de actas contabilizadas por el PREP.

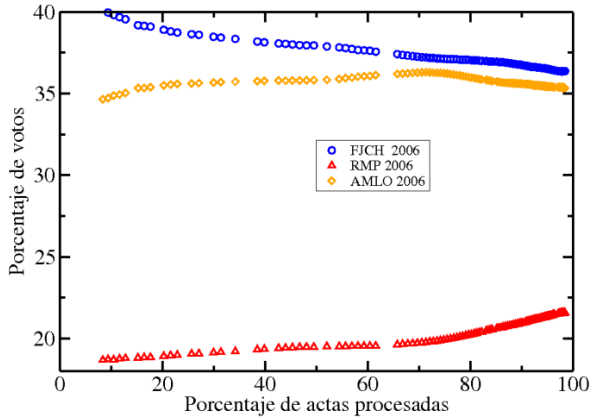

Fig. 1 Porcentaje de votos para los principales candidatos presidenciales en la elección del 2006 como función del porcentaje de actas procesadas. Datos publicados en la página del IFE y capturados por [44].

Se presenta el porcentaje de actas en en el eje horizontal en vez del tiempo, por ser una variable más sencilla de manejar y comparar con otras elecciones. También se ha decidido utilizar esta variable debido a que en el país se han realizado procesos de redistritación los cuales pueden generar dinámicas diferentes de un proceso a otro.

Al dejar de lado la variable tiempo y analizar los datos normalizados, se observa que el esperado cruce de los porcentajes de votos entre los dos punteros nunca se dió. En la madrugada de ese día lunes un cruce evitado hizo sospechar de un fraude electoral. En la Fig. 2 presentamos un acercamiento de los dos candidatos principales que corrobora que efectivamente nunca existe un cruce. ¿Por qué tanto en la Fig. 1 como en la Fig. 2 no se refleja el citado cruce?. Un estudio más cuidadoso del comportamiento de los porcentajes y de otras elecciones ayuda a entender, al menos en parte, este "hecho sospechoso". Los datos iniciaron su publicación después de tener un cierto número de actas capturadas. Esto es alrededor del 10%. La aparición de un salto en los datos ocurre cuando se ha capturado alrededor del 65%. Este comportamiento fue discutido en

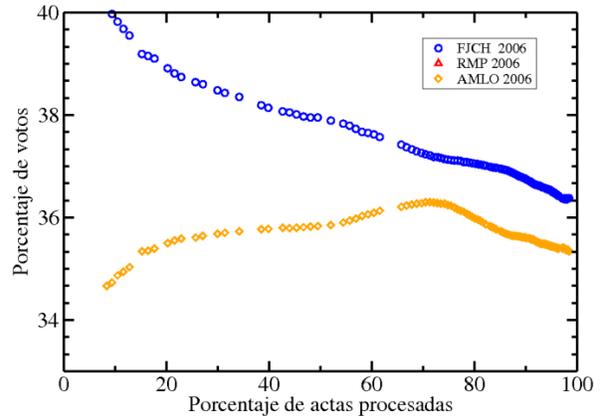

[29,46].

Fig. 2 Acercamiento de la Fig. 1 para mostrar la relación entre los dos primeros candidatos y el cruce evitado

En relación a lo anterior podemos ver un efecto de espejo donde aparentemente uno de los contrincantes le está pasando sus votos al otro (Fig. 2). Este efecto de espejo, es muy común en variables normalizadas. Dado que la norma restringe la suma de porcentajes de voto, si una de las variables se incrementa, la otra, *a fortiori*, tiene que disminuir. Este efecto es fuerte en los datos presentados debido a que los porcentajes de los partidos más pequeños son más o menos constante. Esta cuestión es fácil de entender si se consideran sólo dos partidos con un número de votos al tiempo $t$ igual a $V1(t)$ y $V2(t)$. En este caso, el porcentaje de votos del partido 1 será:

$$\%V1(t) = [V1(t)/(V1(t)+V2(t))] \times 100. \quad (1)$$

De aquí es evidente que el porcentaje de votos de un partido depende del número de votos del otro partido. De hecho, depende del *porcentaje* de votos del otro partido, pues la ecuación anterior la podemos reescribir como

$$\%V1(t) = [(V1(t)/V2(t)) \times \%V2(t)] \times 100. \quad (2)$$

Por ello, es muy natural la existencia de correlaciones y efectos de espejo en cantidades normalizadas, *aún cuando los valores de las variables provengan de un proceso aleatorio*.

Un efecto adicional en la Fig. 1. podemos notar que el tercer contrincante RMP del PRI y el PVEM, tiene un fuerte incremento en el número de votos en las casillas que van llegando en la madrugada de la jornada electoral. Este incremento tampoco es sospechoso pues históricamente ha aparecido en todas las elecciones presidenciales lo cual implica que hay una correlación entre casillas que llegan al final de la jornada y votos para el PRI [26]. Una manera de explicar este comportamiento es considerar que los reportes de las casillas que llegan al último corresponden a sectores del país con un alto índice de marginalidad. Esto ha sido mostrado en parte en [34] y la relación del PRI con altos niveles de marginalidad ha sido bien documentado [45]. Este comportamiento es tan regular que es una de las predicciones cumplidas para la contienda presidencial en 2012 y puede consultarse en [26].

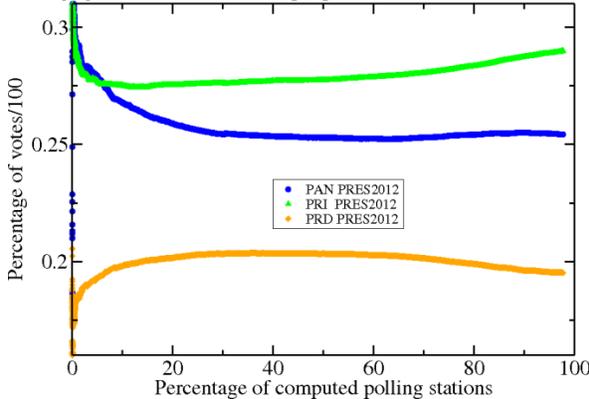

Fig. 3. Porcentaje de votos para los tres principales candidatos en la elección de 2012 como función del porcentaje de actas computadas.

A continuación presentamos otro ejercicio donde se vincula esta idea para dos elecciones presidenciales: la de 2006 y la de 2012.

En este caso utilizamos la relación de los niveles de participación para cada uno de los partidos y su relación con las secciones electorales de las que proceden parecen ser una constante. La Fig. 3 muestra los porcentajes de votos como función del número de actas computadas de las elecciones presidenciales en 2012 para los tres primeros contendientes tomando como referencia el partido mayoritario que representaba cada candidato (PAN, PRI o PRD). La Fig. correspondiente para la elección del año 2000 se encuentra en [24]. En la Fig. 4 reproducimos los mismos resultados de la Fig. 3, pero *artificialmente* añadimos un porcentaje de votos a los candidatos para hacerlos similares a los porcentajes obtenidos en la contienda del 2006. Como podemos ver, los crecimientos y decrecimientos de los porcentajes están correlacionados positivamente. Note que la Fig. 1 y la Fig. 4 ahora son muy parecidas. El ejercicio para el caso presidencial del año 2000 arroja los mismos resultados. Por supuesto que esto no demuestra la aseveración de que los partidos mantienen una misma proporción de electores en las mismas zonas geográficas, pero nos sirve para sostener que el procesamiento de la información sigue un patrón similar independientemente del proceso electoral. Si asumiéramos que la relación entre porcentaje de voto y los índices de marginalidad en el país sea cierta, la Fig. 4 indicaría que los índices NO han cambiado significativamente en los últimos 12 años. En términos de alternancia política, esto implica que la pobreza (y su combate) no ha generado un cambio significativo en el cambio de preferencias políticas para el caso de la elección presidencial.

B. Estadística de radio de participación electoral por municipio.

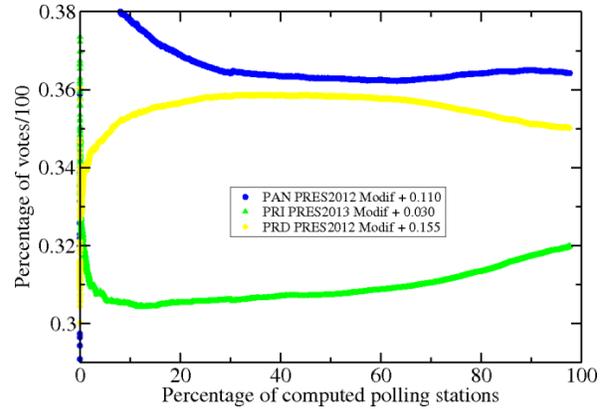

Fig.4 Los resultados de la Fig. 3 pero con los **porcentajes alterados** para simular aquellos del 2006.

Uno de los análisis estadísticos a los que ha sido sometido el caso mexicano es el de la tasa de participación de los votantes (*voter turnout*). En este estudio, debido a C. Borghesi, J.C. Raynal y J.P. Bouchaud [21], se consideraron 77 elecciones en 11 países diferentes: 22 de Francia, 13 de Austria, 11 de Polonia, 7 de Alemania, 5 de Canadá, 4 de España, 4 de Italia, 4 de Rumania, 3 de México, 3 de Suiza y una de la República Checa. En ellas se consideró la variable logarítmica de la tasa de participación, *tp*,

$$tp = log(Na+/Na--) . \qquad (3)$$

Aquí *Na+* es el número de votantes que acudieron a la elección en la municipalidad número *a* y *Na--* es el número de los que no votaron. Por supuesto que $Na = Na+ + Na--$. La distribución de probabilidad de la variable, propiamente desplazada a promedio cero y escalada a varianza uno, es considerablemente similar y estable en todos los casos estudiados. Sin embargo, los autores hacen notar que dicha distribución *no* es gaussiana. El estudio va más allá de plantear que los procesos electorales tienen una distribución similar y logran establecer que existe una dependencia logarítmica de las correlaciones espaciales para las tasas de participación electoral en una amplia gama de elecciones, incluidas las mexicanas. Los resultados mexicanos analizados fueron las elecciones de diputados federales de 2003 y 2009. Lo remarcable, en el contexto de la presente contribución, es que el comportamiento estadístico *no* se diferencía, fundamentalmente, de elecciones francesas, austriacas o polacas. Dicho de otra forma: las elecciones mexicanas se parecen a elecciones de otros países en el mundo, incluyendo casos de democracias consolidadas como la francesa. Las elecciones que muestran mayores desviaciones son las alemanas.

Como dato interesante y asumiendo la validez de la regularidad entre elecciones indicada por estos autores la misma tasa de participación y su distribución en varios países, incluído México ha sido usada para sospechar del éxito del partido que soporta al Presidente de Rusia, Vladimir Putin, en las elecciones a la Duma en 2011[47]. Aunque no lo hemos verificado, la distribución con la que se compara es la elección a Diputados en México en 2009. Nuevamente, los datos electorales mexicanos muestran consistencia con los datos de elecciones europeas, en este caso Polonia, primera vuelta presidencial del 2010; Bulgaria, elección parlamentaria de 2009; Ucrania, primera ronda presidencial 2010.

Como una conclusión a esta sección podemos notar que algunas de las propiedades estadísticas reportadas en el caso mexicano concuerdan con el comportamiento en elecciones europeas

### C Benford a Segundo dígito no se cumple en las elecciones mexicanas.

Una de las técnicas que se han propuesto para detectar anomalías en los resultados de una elección es la ley de Benford a segundo dígito significativo [35]. Sobre los problemas en usar el primer dígito significativo referimos al lector a [48] y las referencias en ella. En el caso mexicano esta prueba ha sido utilizada por V. Romero para fundamentar su argumento de fraude en las elecciones presidenciales del 2006. Una explicación al respecto se encuentra en [30].

Por cuestiones de espacio no nos extenderemos en este tema pero en el último reporte de Mebane[35], se concluye "... Using excerpts from my book manuscript Election Forensics, which examine data from elections in the United States, Germany, Canada and Mexico, I assess the performance of tests based on the second significant digits of precinct-level vote counts. The claim that deviations in vote counts' digits from *the distribution implied by Benford's Law is an indicator for election fraud generally fails*." (Las cursivas son nuestras).

### D. Agentes en la detección del fraude en 2006

En [28] se aplica el modelaje con agentes como una técnica novedosa para la detección de fraudes electorales, ello, de nuevo, a partir del análisis estadístico de los resultados. En este modelo los agentes computacionales dan resultados dependiendo de sus interacciones con la sociedad y de información global. Este caso corresponde al modelo de elecciones limpias. Para simular el efecto fraudulento se manipula a un cierto porcentaje de agentes (votantes) y se pretende descubrir con métodos estadísticos si la elección ha sido manipulada o no.

El modelo es calibrado para las elecciones del año 2006 para la presidencia de México con datos pre- electorales. Bajo estas condiciones los autores de [28] concluyen que "*el modelo de simulación descarta la presencia de un fraude de gran magnitud en el que no menos del 5-6% del total de votos emitidos son manipulados.*"

### E. Distribución de Errores y sesgos en favor de un partido o candidato.

Muchos de los procesos con los que tratan los físicos tienen un naturaleza estadística, existe una variación en el valor final obtenido en cada uno de los experimentos realizados y dependen de la muestra considerada dentro de la población[41]. El carácter probabilístico de la mecánica cuántica, por ejemplo, es que en el fondo no podemos obtener una respuesta única al realizar un experimento. Tenemos que realizar un conjunto de mediciones en los que obtenemos diferentes resultados. De esos resultados podemos calcular los valores esperados, promedio, con una cierta variación y eso se manifiesta en los postulados de la mecánica cuántica. Estos postulados son de naturaleza probabilística, ver por ejemplo [49]. En el caso electoral los resultados dependen de la muestra y de una serie de factores que alteran el resultado final. Si la medición se encuentra entre el valor A y B, existen una serie de consideraciones. Una cuestión obvia es si estamos considerando a TODA la población. En el caso de las elecciones presidenciales en México a partir del 2006 se habla de la existencia de sesgos sistemáticos en contra de un partido o candidato llegando incluso a ser un argumento para la anulación de la elección de 2006. ¿Hasta dónde es posible que la distribución de errores refleje un sesgo intencional en favor de un partido o candidato?. Presentamos algunos argumentos para refutar esta idea. Primero, no todos los electores que están inscritos en el padrón, fueron a votar el día de la elección. Existió un porcentaje de ellos que no expresaron su opinión. Segundo, existe una incertidumbre en el número de votantes que quisieron ejercer su derecho pero que, por estar fuera de sus localidades, por ejemplo, no lo pudieron expresar. La autoridades electorales han colocado casillas especiales en varios lugares para los votantes en tránsito. Sin embargo, éstas tienen un límite de papeletas electorales por lo que a pesar de estar distribuidas a lo largo del país resultaron insuficientes para la gran cantidad de electores que llegaron ese domingo de julio de 2006. Muchos de ellos quedaron sin boleta por llegar tarde o simplemente por llegar más tarde que los votantes duros de los diferentes partidos. Por supuesto, para muchos estudios es suficiente con considerar la población total como aquella que fue a votar y dicho voto fue registrado.

Aunado a los problemas anteriores de las casillas electorales especiales, los errores en el conteo han sido generales, en lo que se puede denominar errores aritméticos en el conteo del PREP. El Programa contabiliza, además del número de votos para cada uno de los candidatos, el número de boletas sobrantes, el número de votantes que asistieron, número de boletas recibidas y depositadas. El conjunto de errores de consistencia que pueden formarse son seis y son de la forma: *número de boletas depositas debe ser igual al número total de votos para cada partido, más los votos nulos y los votos para candidatos independientes*. Aritméticamente la diferencia de estas cantidades debe ser cero. Si esto ocurre para todas las casillas la distribución de esta variable es una delta de Dirac. Ya en [31,32] se comentó al respecto, sin embargo la distribución de errores va más allá de lo comentado por el autor de esos trabajos. Como ha sido explicado en [26], los errores en los campos de control que aparecen en los registros del PREP tienen características generales tan regulares que han permitido ser predichas para las recientes elecciones presidenciales de julio de 2012.

En la Referencia [31] se comenta que "los errores aritméticos no son parte del error muestral que desaparezca al acumular más y más casillas, sino que constituyen un error de medición al contar boletas y asentar los resultados en actas" y ello está presente en los procesos de 2006 y 2000 donde " las actas de elección presidencial de 2006 tuvieron menos errores (46.7% del total) y menos datos omitidos que las de 2000 (51.4%). Sin embargo, la magnitud promedio de esos errores es muy similar... En ambas elecciones, los errores aritméticos se distribuyeron de manera casi uniforme o aleatoria en las casillas ganadas por uno u otro candidato, lo cual sugiere errores humanos aleatorios que no reflejan intencionalidad de cometer fraude hacia un partido u otro."

Los resultados anteriores han sido ampliados no solo para los valores porcentuales, sino que se han obtenido las distribuciones de error descritos arriba con una remarcable similitud en su estructura [26]. En todas ellas el centro de la distribución aparece como una ley de potencia con lóbulos a los lados. Y ello ocurre para tres elecciones presidenciales en México, 2000, 2006 y 2012. La primera no ha sido cuestionada sobre acciones fraudulentas del ganador o del perdedor aún cuando tiene una importancia histórica mayor, pues el partido en el poder perdió esa elección después de siete décadas. La de 2006 fue marcada por acusaciones de fraude y la última, marca el regreso del PRI después de una campaña de confrontación exacerbada y la existencia de un movimiento estudiantil en contra abiertamente del candidato del PRI. En estos tres momentos con condiciones políticamente diferentes las distribuciones son parecidas, tienen el mismo comportamiento general. Esto sugiere la presencia de dinámicas similares en los tres eventos pese a las diferencias contextuales. Un claro entendimiento del origen de estos errores es necesario y se encuentra en este momento bajo estudio por nosotros.

### F. Leyes de Potencia

Otro aspecto polémico de los análisis electorales es la existencia o no de leyes de potencia en algunas de las cantidades medidas. En general, la aparición de este tipo de distribuciones señala que no existen escalas predominantes en los fenómenos. La existencia de leyes de potencia en la naturaleza es de por si materia de amplio debate pues es un problema delicado y en muchas ocasiones se describen como leyes de potencia fenómenos que no lo cumplen. Una excelente referencia para entender esta problemática se encuentra en [50] donde de manera cuidadosa se discute el tema.

Las leyes de potencia en datos empíricos han aparecido en la distribución de la fracción de votos que obtienen los candidatos o partidos contra el número total de votos expresados en la sección o distrito. La existencia de muchos candidatos para las posiciones en las diferentes cámaras o para las municipalidades hace que esa fracción de votos sea una variable estadística.

Se puede considerar para su estudio la distribución de esta variable o la versión acumulativa de ella. En cualquiera de los casos la potencia, $v$, con la que decaen parece depender del país y del tipo de proceso. Para las elecciones brasileñas a diputados estatales y federales de 1998 y 2003 la potencia es $v = 1.00$ [6,7]. Para las elecciones en India durante 1998, en [10] se reportan valores variados para las diferentes regiones, desde $v = 0.97$ para las elecciones en Goa hasta 2.06 en Haryana. El valor nacional está en 1.30. Las elecciones Indonesias de 1999 y 2004 los valores reportados tampoco son uniformes [52]. En [11] el ajuste va un poco más allá de ser una ley de potencias simple y llana analizando un ajuste de ley de potencia con un corrimiento conocida como ley de Zipf y una ley de Zipf generalizada que proporciona un decaimiento en ley de potencia para algunas décadas y una región curvada para otras escalas. Estos modelos logran hacer un ajuste a las diferentes dinámicas que los electores pueden tener. En [12] encuentran un modelo que reproduce el decaimiento como una log-normal para varias elecciones europeas

En [23], se hace el análisis para las elecciones legislativas en México en 1991,1994, 1997, 2000 y 2003. En ellas se eligen 300 diputados para la cámara baja federal por elección directa. Se presentan candidatos por una gran variedad de partidos nacionales y locales en los diferentes distritos electorales. Los resultados obtenidos son de $v = 1.23$ para candidatos y 1.444 para partidos. Todos los valores fueron obtenidos para la distribución acumulada. No obstante el decaimiento solo es aparente en una década.

Pese a estos resultados, la investigación sobre estos temas continúa. La falta de universalidad en estas distribuciones puede ser resultado de varios factores, entre ellos que el número de electores en cada sección o distrito vote por un candidato particular, del grado de participación política electoral en la sección particular y del carisma particular de algunos de los candidatos dentro de sus secciones. Esto último es claro cuando algunos candidatos tienen particular incidencia en los problemas locales y atraen votos de personas que no coinciden con el programa de su partido, pero sí con el actuar del candidato. Recientemente, en [51], se estudió en las elecciones brasileñas el efecto del número de miembros del partido y el número de candidatos a alcalde y consejero. En ese estudio se encontró que existe una relación tipo ley de potencia rodeada de ruido log-normal multiplicativo entre el número afiliaciones al partido y la población de votantes. Esto es diferente a lo que este grupo encontró para número de candidatos y población de votantes. Parece que la importancia del cargo da lugar a estadísticas y dinámicas diferentes.

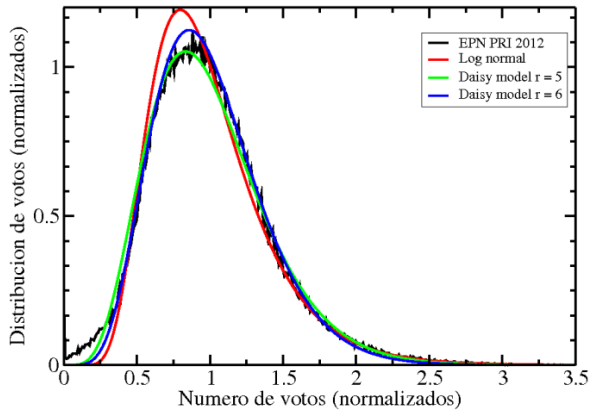

Fig.5 Distribución de voto para el PRI en la elección presidencial del 2012. Se presentan algunos posibles ajustes teóricos.

G. Distribución de voto a favor de partidos.

Además del nivel de participación uno podría preguntarse por la distribución de voto para cada casilla, esto es: en cuantas casillas un partido obtiene cero votos, un voto, dos votos y así. El histograma resultante puede normalizarse para que el área bajo la curva sea uno, pero además, la variable misma ha de normalizarse de una manera particular. Si el promedio de votos es más o menos el mismo no importando que subconjunto de datos tomemos, podemos normalizar la variable a promedio uno simplemente dividiendo los valores por el valor promedio. Si esto no es así el procedimiento es más delicado. Para los casos que discutimos aquí no hay tal complicación. En [25-27] se encontró que la distribución de voto debidamente normalizada para el PRI era posible hacer el ajuste de una distribución de probabilidad conocida como modelo margarita. Su forma particular se puede consultar en las referencias. Aquí mostramos un ejemplo para la votación del PRI para presidente en 2012 (Fig. 5). Ahí se compara la distribución empírica con uno de los modelos margarita de rangos 5 y 6, además del mejor ajuste de una distribución log-normal. Como puede observarse, el ajuste es delicado, pero los modelos margarita dan un buen ajuste. Se puede hacer un ajuste a un distribución Gamma, que depende de dos parámetros con un buen ajuste. En la Fig. 6 se muestran los mismos casos pero en escala semi-log. Por supuesto que la validación de la hipótesis de que alguna de estas distribuciones modele los datos experimentales requiere de un test como el de Kolmogorov-Smirnov, pero para la discusión que sigue no es relevante.

En las referencias [25-27] se ha discutido la distribución

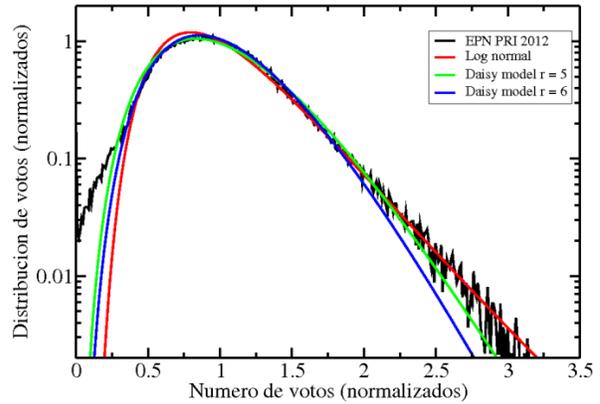

del PRI en elecciones presidenciales del 2000 al 2012 y en

Fig. 6 Mismas distribuciones que la Fig. anterior, pero en semi-log.

elecciones para ambas cámaras en diferentes periodos. En todos ellos la distribución presenta una estructura general muy parecida a la descrita aquí. Supongamos que deseamos comprobar que en las elecciones del 2012 hubo una compra masiva de votos. ¿Qué clase de aproximación podemos usar para descubrir esa compra? Como hemos discutido en la sección C la ley de Benford no es confiable para detectar actividades fraudulentas, en la sección D se discutió que si la actividad fraudulenta no es suficientemente masiva no es posible detectarla. *¿Podríamos usar aproximaciones de tipo entropía de Shannon para descubrir la manipulación de votos?* Para una elección dada podemos tratar de comparar el comportamiento de la entropía de alguna distribución de probabilidad de los partidos. La esperanza es que el comportamiento general de la entropía fuera cualitativamente similar no importando de que partido se trate. Así, si alguno de los partidos ha realizado compra de votos, su entropía cambiaría. El problema con esta aproximación es que no importando que variable seleccionemos, tenemos que conocer el comportamiento entrópico de los partidos en varias elecciones anteriores en las que no hubiera sospecha de fraude. O alternativamente, proponer un modelo, como el de agentes, para calibrar una elección confiable que sirva de contrafactual para comparar las elecciones fraudulentas. Todo esto en el mejor espíritu de la prueba de hipótesis planteada en la sección II.C. Ahora bien, el caso es que conocemos, al menos en parte, el comportamiento del PRI sospechoso de comprar votos en 2012 a lo largo de al menos dos situaciones políticamente diferentes. Así que es de esperarse que cualquier distribución de probabilidad que involucre de alguna manera

la distribución de votos NO PODRÁ ser cualitativamente diferente en la elección de 2012 que en cualquier otra. Por lo tanto, las medidas basadas en la entropía de Shannon que use alguna variación de la distribución de voto será incapaz de descubrir alguna anomalía en las elecciones presidenciales del 2012 en México.

IV- Discusión: Fraude o no fraude

Toda la evidencia que hemos presentado en esta contribución haría pensar que no hay fraude o alteración en las elecciones mexicanas, pues sigue de manera bastante general las propiedades estadísticas encontradas en otras elecciones nacionales o de democracias consolidadas. Sin embargo, para el caso de México ha sido bien documentado el hecho de compra de votos por parte de todos los partidos en las elecciones de las últimas dos décadas [53], la existencia de operadores políticos que garantizan el voto duro y toda una pléyade de maneras de corromper el voto ciudadano.

Sin embargo y de manera preocupante, estos hechos no son privativos de las elecciones mexicanas y también han sido ampliamente documentados en elecciones italianas y brasileñas. Seguramente existe evidencia de su existencia en muchas otras elecciones y países. Lo que nos indica la evidencia presentada aquí es que en un proceso electoral existen una serie de regularidades dentro de las cuales se podría incluir desafortunadamente la manipulación del voto. También nos indica que es menester afinar nuestras herramientas para entender el origen de éstas regularidades (estadísticas) y poder, eventualmente, entender la dinámica de fraude. La naturaleza ha tenido millones de años para probar y desechar sistemas biológicos, pero también sociales [54,55]. ¿Quién nos dice que nuestras burdas herramientas pueden analizar dichos comportamientos? La física ha analizado solo la parte fácil en la naturaleza hasta el momento. Queda por ver si puede arrojar alguna luz sobre los procesos, por demás complejos, de las sociedades humanas.

V. A manera de Conclusión.

Lo más importante es que la discusión se ha abierto y que más allá del fraude o no en las elecciones presidenciales de 2006 en México, el estudio de sistemas electorales con el espíritu y lsa herramientas de la física está en progreso.

REFERENCIAS

[1] Ph. Ball."The Physical Modeling of Human Social Systems", Complexus, 1 (2003) 190.
[2] Ph. Ball Ball, Philip. (2004). Critical mass - How One Thing Leads to Another ("particles become people", pg. 110). New York: Farrar, Straus and Giroux. Hay version al espannol en *mása crítica*. Cambio, caos y complejidad (FCE-Turner, 2010)
[3] Z. Neda, E. Ravasz, Y. Brechet, T. Vicsek and A-L Barabasi. "The sound of many hands clapping"Nature (London) 403 (2000) 849
[4] D. Helbing, I. Farkas and T. Vicsek, Nature (London) 407 (2000) 487
[5] D. Helbing and M. Treiber, Science (1998) 2001
[6] R.N Costa Filho, M.P. Almeida, J.S. Andrade Jr, J.E. Moreira "Scaling behavior in a proportional voting process" Phys Rev E 60 (1999) 1067.
[7] R.N Costa Filho, M.P. Almeida, J.S. Andrade Jr, J.E. Moreira "" Physica A 322 (2003) 698.
[8] A.T. Bernardes, D. Stauffer, J. Kertesz. Eur. Phys. J. B. 25(2002) 123.
[9] G. Treviso, L.F. Costa. Phys. Rev. E. 74 (2006) 036112.
[10] M.C. Gonzalez, A.O. Sousa, H.J. Herrman, Int.J. Mod. Phys C. 16 (2004) 45
[11] Lyra. M.L. U.M.S. Acosta, R.N Costa Filho, J.S. Andrade Jr "Generalized Zipf's law in proportional voting process". Europhysics Lett. 62(2003) 131.
[12] S. Fortunato, C. Castellano, Phys. Rev. Lett. 99 (2007)138701
[13] L.E. Araripe, R.N. Costa Filho, Physica A 388 (2009) 4167
[14] A. Chatrerjee, M. Mitrovic, S. Fortunato, Sci. Rep. 3 (2013) 1049
[15] L.E. Araripe, R.N. Costa Filho, H.J. Herrmann, J.S. Andrade, Int. J. Mod. Phys C 17 (2006)1809
[16] N.A.M. Araujo , J.S. Andrade, H.J. Herrmann, "Tactical Voting in Plurality Elections" PloS ONE 5 (2010) e36289
[17] M. Tumminello, S. Micciche, J. Varho, J. Piilo, R. N. Mantegna,"Quantitative Analysis of Gender Stereotypes anInformation Aggregation in a National Election" PLoS ONE 8, e58910 (2013)
[18] M. C. Mantovani, H. V. Ribeiro, M. V. Moro, S. Picoli Jr., R. S. Mendes, EPL 96, 48001 (2011)
[19] P. Klimek, Y. Yegorov, R. Hanel, S. Thurner, PNAS 109, 16469 (2012)
[20] C. Borghesi, J.-P. Bouchaud, Eur. Phys. J. B 75, 395 (2010)
[21] C. Borghesi, J.-C. Raynal, J.-P. Bouchaud, PloS ONE 7, e36289 (2012)
[22] C. Borghesi., J. Chiche, J.-P. Nadal PloS ONE 7, e39916 (2012)
[23] O. Morales-Matamoros, M.A. Martinez-Cruz, R. Tejeida-Padilla. "Mexican voter network as a dynamical complex system"
[24] G. Báez, H. Hernández-Saldaña, R.A. Méndez-Sánchez. "On the reliability of voting processes: the mexican case". Arxiv 13/jun2012 060911v3. Primera version Sept. 2006.
[25] H. Hernández-Saldaña. `` On the corporate votes and their relation with Daisy Models". Physica A 388 (2009) 2699-2704.
[26] H. Hernández-Saldaña, M. G\'omez-Quezada and E. Hern\'andez-Zapata,``Statistical distributions of quasi-optimal paths in the traveling salesman problem:The role of initial distribution of cities". Rev. Mex. F\'is. S. 58(1) (2012) 0032.
[27] H. Hernández-Saldaña.``Results on ``Three predictions on July 2012 Federal election in México based on past regularities" "{arXiv:1212.6676v1 [physics.soc-ph] 30 Dec 2012.
[28] G. Castañeda, I. Ibarra. "Detección de fraude con modelos basados en agentes: las elecciones mexicanas de 2006" Perfiles Latinoamericanos 36. Julio -diciembre 2010. 43
[29] W.L. Mochan. http://em.fis.unam.mx/blog
[30] H. Diaz-Polanco. "La cocina del Diablo. El fraude de 2006 y los intelectuales" México. Temás de hoy. 2012. 247p.
J. Aparicio. "análisis estadistico de la elección presidencial de 2006". Politica y gobierno. Volumen tematico 2009. Eleciones en México. 225-243.
[31] J. Aparicio "La evidencia de una elección confiable" Nexos. Num 346. Octubre de 2006
[32] J. A. Crespo. "2006: hablan las actas. Las debilidades de la autoridad electoral mexican". México. Debate. 2008. 232p.
[33] F. Pliego Carrasco. "El mito del fraude electoral en México". México. Editorial Pax México. 2007. 218p.
W. R. Mebane Jr."Election Forensics: The Meanings of Precinct Vote Counts' Second Digits" presentation at the 2013 Summer Meeting of the Political Methodology Society,University of Virginia, July 18–20, 2013. http://pages.shanti.virginia.edu/PolMeth/files/2013/07/Mebane.pdf
[34] http://www.ife.org.mx/portal/site/ifev2/estadísticas_y_Resultados_Electorales/
[35] http://www.ife.org.mx/portal/site/ifev2/Programa_de_Resultados_Electorales_Preliminares_PREP/
[36] https://ciudadania.ife.org.mx/infomex/ActionInitSAILoginINFOMEX.do

[37] PANDAS http://pandas.pydata.org/ Phyton Dat Analysis Library. Consultado 17 Agosto 2013.
  The R Project for Statistical Computing http://www.r-project.org/
[38] P.I. Good and J.W. Hardin. "Common Errors in Statistics (and how to avoid them)" 3[rd]. ed. USA. Wiley. 2009.
[39] P. Cummings and T.D.Koespsell, "Statistical and design issues in studies of groups" Inj. Prev. 8 (2002) 6-7
[40] L.C Ugalde. "Así lo viví Testimonio de la elección presidencial de 2006, la más competida en la historia moderna de México". México. Grijalvo. 2008
[41] Baqueiro. Comunicacion personal.
[42] A. Moreno. "El votante mexicano: Democracia, actitudes políticas y conducta electoral". México DF: Fondo de Cultura Económica. 2003 252 p.
[43] W.L. Mochan. "Incertidumbre y errores en las elecciones de julio de 2006. " Ciencias 84 octubre noviembre (2006) 39
[44] S. Shpilkin. "Elections improbability". Gazeta.ru. 13.12.2011 translated by S.. Kvasha. Consultado el 20.12.2011 11:40 am http://en.gazeta.ru/news/2011/12/13/a_3926402.shtml.
[45] W.R.Mebane, Jr, "Election Forensics: Statistics, Recounts and Fraud" Annual Meeting of the Midwest Political Science Association, Chicago,IL,USA. 2007.
[46] C. Cohen-Tannoudji, B. Din, F. Laloe. Wiley Interscience, N.Y., USA. 1977 (2 vols.).
[47] M.J.E. Newman "*Power laws*, *Pareto distributions* and Zipf's lawower laws". ArXiv/cond-mat/0412004 .2004
[48] M. C. Mantovani, H. V. Ribeiro, M. V. Moro, S. Picoli Jr., R. S. Mendes. "Engagement in the electoral process: scaling laws and the role of the political positions". ArXiv: 1308.2857v1 [physics.soc-ph] 13 Aug 2013.
[49] H. Situngkir, "The power law signature in Indonesian Legislative election". ArXiv:nlin.AO/0405002. 2004.
[50] [52a] A. Diaz-Cayeros, B. Magaloni,"The politics of public spending. Part I--The logic of vote buying", Background Paper for World Development Report 2004, 2003.
[51] F. De Waals."Chimpanzee Politics: Power and Sex among Apes".The Johns Hopkins University Press; 25th anniversary edition (August 30, 2007)
[52] E.O. Wilson "Sociobiology: The New Synthesis". Belknap Press of Harvard University Press; Twenty-Fifth Anniversary Edition edition (March 4, 2000)